\documentstyle[12pt,twoside,fleqn,epsfig,rotate,espcrc1]{article}
\def\bge{\begin{equation}}
\def\ene{\end{equation}}
\newcommand{\nc}{\newcommand}
\nc{\non}{\nonumber}
\def\be{\begin{equation}}
\def\ee{\end{equation}}
\def\bga{\begin{eqnarray}}
\def\ena{\end{eqnarray}}
\def\eea{\end{eqnarray}}
\def\bg{\begin{eqnarray}}
\def\en{\end{eqnarray}}

\def\ra{\rightarrow}

\newcommand{\AmS}{{\protect\the\textfont2
  A\kern-.1667em\lower.5ex\hbox{M}\kern-.125emS}}

\title{Hadron structure after 25 years of QCD} 

\author{A. W. Thomas\address{
   Department of Physics and Mathematical Physics and \break 
   Special Research Centre for the Subatomic Structure of Matter,\break
   University of Adelaide, Australia 5005}}
\begin{document}
\def\bra#1{{\langle #1{\left| \right.}}}
\def\ket#1{{{\left.\right|} #1\rangle}}
\def\bfgreek#1{ \mbox{\boldmath$#1$}}
\maketitle
\vspace{-9.9cm}
\begin{flushright}
{\footnotesize Invited talk presented at Few Body XVI} \\
{\footnotesize Taipei, Taiwan. March 6-10, 2000} \\
{\footnotesize ADP-00-35/T418 \hspace{2cm}}
\end{flushright}
\vspace{8.9cm}
%
%
\begin{abstract}
We briefly review the status of our understanding of hadron structure based
on QCD. This includes the role of symmetries, especially chiral
symmetry, and the insights provided by lattice QCD. The main focus is on
baryon structure and especially the nucleon, but this cannot
be treated realistically without reference to spectroscopy.
Our aim is to highlight recent insights and promising
directions for future work.
\end{abstract}

\section{INTRODUCTION}

In the space allotted it is impossible to do justice to more than 40
years of work on hadron structure. Rather than attempt this, we take very
seriously the appearance of QCD in the title and concentrate on the
guidance that recent developments in our understanding of QCD provide
for future developments in modelling hadron structure. For reviews of
the vast amount of existing information on hadron models we refer to
recent conferences on hadron structure.

In order to place hadron models firmly in the context of QCD, we
begin with a summary of the properties of the non-perturbative vacuum, 
upon which everything else is built. We then recall some lessons from
heavy quark systems before turning to the more complicated case of
light quarks, where chiral symmetry plays a key role. We finish with a
brief outlook concerning the theoretical and experimental
possibilities in the next decade.

\section{THE QCD VACUUM}
  
It is by now firmly established that the ground state of QCD is a highly
non-trivial state, including both quark and gluon condensates. For a
purely gluonic version of QCD the vacuum energy density, $\epsilon_{\rm
vac}$, is:
\be
\epsilon_{\rm vac} = - \frac{9}{32} \langle 0 | \frac{\alpha_s}{\pi}
G^2 | 0 \rangle = - 0.5 {\rm GeV/fm}^3.
\label{eq:1}
\ee
In comparison with phenomenological estimates of the 
energy difference between the
perturbative and non-perturbative vacuum states, such as $B$ in the MIT
bag model, this is an order of magnitude larger. Thus either the simple
idea of the perturbative vacuum being fully restored inside a hadron is
incorrect or the situation is rather more complicated than usually
assumed.

The light quark condensates, $\langle \bar u u \rangle$ and $\langle
\bar d d \rangle$, in the non-perturbative vacuum are approximately
equal and take a value around (-240 MeV)$^3$. A quantitative
understanding of these values requires a treatment that in nuclear
physics terms would involve at least Hartree-Fock plus RPA
\cite{cotanch}. The underlying chiral symmetry of QCD (for massless $u$
and $d$ quarks) requires that, if
the vacuum is to have a non-trivial quark condensate, there must be 
massless Goldstone bosons with the quantum numbers of the pion. As we
move away from the massless limit, these Goldstone bosons have masses
which behave as:
\be
m_\pi^2 \propto \frac{\bar m_u + \bar m_d}{2} = \bar m.
\label{eq:2}
\ee
(In fact, while this relation is only formally correct near $\bar m =
0$, lattice simulations show that it holds for $m_\pi$ as large as 1
GeV.)

By studying the energy of two static sources of opposite color charge on
a lattice, it has been shown that purely gluonic QCD leads to a linearly
rising potential at large separations, $V(R) = \sigma R$. This
observation naturally explains the observed Regge
trajectories and effectively ``confines'' heavy color sources.
Supplemented by a  short-range, one-gluon exchange potential this
naturally yields a potential of the Cornell type, which has
proven phenomenologically so successful for heavy quark-anti-quark
pairs:
\be
V_{Q \bar Q}(r) = C - \frac{\alpha}{r} + \sigma r.
\label{eq:3}
\ee

A linearly rising potential at large separations can be understood in
terms of a non-perturbative vacuum which has dia-electric properties. By
analogy with the Meissner effect in superconductors, one can think of
the QCD vacuum containing a condensate of color magnetic monopoles which
shield the lines of color electric force into a tube of constant cross
section, no matter how far apart the color sources are located. Such a
picture is supported by lattice simulations involving Abelian
projection, but the final details are not yet settled \cite{Toki}.
On the other hand, in the real world this situation is altered
dramatically by the presence of light quarks which can break the string
once the energy stored becomes too great, through the process $Q \bar Q
\ra Q \bar q + \bar Q q$ -- e.g., for charm quarks, $c \bar c \ra D \bar
D$. For mesons which are stable under strong interactions, virtual
processes like this have a quantitative effect -- e.g., $f_B = 220$ MeV
in quenched QCD (QQCD), whereas it is 260 MeV in full QCD -- 
but do not change the qualitative picture \cite{bernard}.

\subsection{Insights from lattice QCD}

\begin{figure}[tbp]
\begin{center}
\epsfig{figure=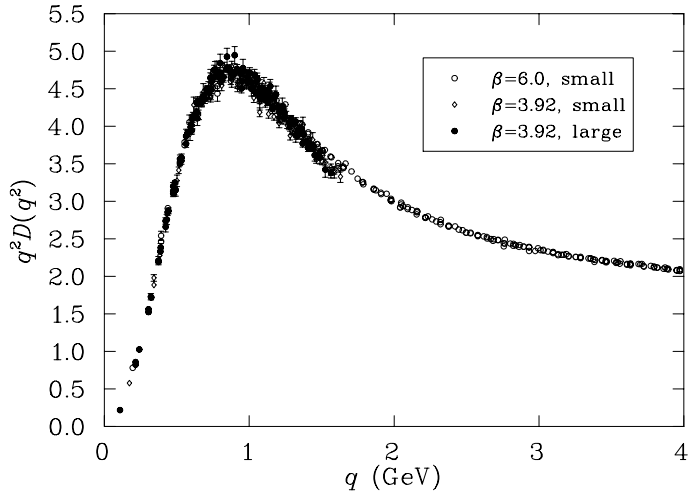,angle=90,width=9cm}
\end{center}
\caption{
Non-perturbative behaviour of the gluon propagator (times $q^2$) 
in Landau gauge, 
calculated from lattice QCD -- from Ref. \protect\cite{gluonprop}. }
\label{fig:AllProps}
\end{figure}
%
%
It has recently proven possible to compare various intermediate stages
of lattice calculations, such as the quark and gluon propagators, with
the forms commonly used in models. In this way one can refine the model
building process using QCD itself. Of course, intermediate steps such as
the quark and gluon propagators are not physical and one must
specifically fix the gauge in order to make a meaningful comparison. The
gauge most commonly used is Landau gauge and techniques have been
developed to fix lattice quantities in this gauge. Figure 1 shows the
result for the non-trivial momentum dependence of the gluon propagator
(times $q^2$) \cite{gluonprop},  
$q^2 D(q^2)$, which should go to a constant at large $q^2$ (up to 
perturbative QCD logs). From Fig. 1 
we see that the lattice simulation shows that the gluon
propagator is clearly {\it non-perturbative} for $q^2 < 4$GeV$^2$.
Even more interesting from the point of view of model building is the
fact that the gluon propagator is not enhanced as $q^2 \ra 0$.
While this agrees with some recent Schwinger-Dyson studies of QCD
\cite{alkofer}, it is in disagreement with at least a naive interpretation
of a great deal of phenomenological work related to dynamical chiral
symmetry breaking within the Schwinger-Dyson formalism 
\cite{roberts}. Clearly this sort of interplay between phenomenological
models and QCD itself has just begun and we have a great deal to learn
from it.

Again with Landau gauge fixing, there have been some preliminary studies of
the quark propagator in QCD. For Euclidean $p^2$ one can write the quark
propagator as:
\be
S_E(p) = \frac{Z(p^2)}{i \gamma^\mu p_\mu + M(p^2)}.
\label{eq:4}
\ee
The lattice simulations, which have so far been carried out with
relatively large current quark masses, show a clear enhancement in the
infrared \cite{quark,aoki}. For example, for 
a current quark mass of order 110 MeV, the
simulations suggest $M(0) \sim 400$ MeV, decreasing to around 300 MeV in
the chiral limit. This is certainly consistent with the general idea of
the constituent quark model and indeed this result provides a firm
theoretical foundation for the concept within QCD. Of course, it also
indicates where the concept breaks down and it is clear that in
processes involving significant momentum transfer it will be necessary
to go beyond the simple idea of a fixed mass. The similarity of the 
mass function, $M(p^2)$, to that found in Schwinger-Dyson studies
\cite{SDE} suggests that the latter may be a promising phenomenological
extension of the constituent quark idea. 

\section{NON-RELATIVISTIC QUARK MODEL}

The non-relativistic quark model, which combines the concept of a
massive constituent quark with a linear, pairwise confining interaction
and a short distance, spin-dependent hyperfine interaction, has had
enormous success in correlating a vast amount of hadronic data. It
provides an excellent basis for understanding the spectrum of mesons and
baryons. Most notably, it has also provided a very natural explanation
of why some states expected in the naive quark model have {\em not} been
seen yet \cite{IK_missing}. While until now this explanation has seemed
convincing, it has not been seriously tested experimentally. This is now
in process of change with Jefferson Lab due to produce a tremendous
quantity of new, high duty factor data.

Amongst the key issues confronting the simple quark model, we mention
relativity, the spin-orbit problem and chiral symmetry. It is quite
clear that the approximation that the constituent $u$,$d$ and $s$ 
quarks are non-relativistic cannot be quantitatively reliable. For
quarks of mass 300--400 MeV confined a volume of radius less than 1 fm,
it is clear that $p/M$ is of order one. The relativistic extension of
the naive quark model, developed by Capstick, Isgur and others 
\cite{capstick} has
overcome this problem, with a resultant improvement in various transition
form factors. 

The spin-orbit problem has been around since the beginnings of the quark
model, but has been of considerable interest recently in the light of
claims that pseudoscalar meson exchange, rather than gluon exchange, as
a source of hyperfine splitting would resolve the problem \cite{Gloz1}. 
On the other hand, as noted
by Isgur \cite{nathan}, 
with a Lorentz scalar confining potential one will automatically have
large spin-orbit forces. The spin-orbit force arising from the usual
one gluon exchange interaction can cancel this in the meson spectrum but
not in the baryon spectrum. Thus the spin-orbit problem for the baryon
spectrum is still very much with us -- see however Ref. \cite{Gloz2}.

As we have already seen in connection with the structure of the
non-perturbative vacuum, chiral symmetry is expected to play a major
role in determining hadron structure for light quarks. Unfortunately the
constituent quark model destroys that symmetry rather badly.  
We shall turn to a specific discussion of the role of chiral symmetry
later, but simply note here that a quantitative discussion of hadronic
properties is not possible unless
measures are taken to ensure that the theory respects chiral symmetry.

\subsection{Hybrids}

In a more sophisticated treatment of hadron spectroscopy the simple
linear confining potential in Eq.(3) may be replaced by a flux tube model. This
has the advantage that the confining potential then becomes truly
dynamical and in particular can be excited. Excitations of the flux tube
can result in a new kind of hadron. The most interesting cases are
those where the quantum numbers of the hadron are {\em exotic} -- 
e.g., for mesons where they cannot be associated with a $q \bar q$
system. 

The experimental discovery of exotic systems, where the gluons
have a genuine structural role, would be a vital step towards a full
understanding of QCD. This explains the excitement over the announcement,
from E852 at Brookhaven National Lab \cite{PRL81_5760}, 
of three candidates for hybrid mesons with quantum numbers $J^{PC} =
1^{-+}$. The $\pi '(1370)$ was seen in the $\pi \eta$ and $\pi \eta '$
channels, the $\pi '(1640)$ in $\pi \eta ', \rho \pi$ and $f_1 \pi$ and
the $\pi '(2000)$ in $a_1 \eta$. These masses are somewhat lower than
the values usually reported in lattice simulations, although for the
moment the latter tend to be based on quenched QCD \cite{Hyb_latt}.
While the interpretation of the BNL data should 
become clearer over the next few years, the announcement lends even
greater urgency to the calls for a future HALL D program at Jefferson
Lab.

\subsection{Glueballs}

An even more dramatic prediction of QCD than exotic states is, of
course, the possibility of physical particles containing {\em only} glue
-- the glueballs. Lattice simulations suggest that the lowest mass state
of pure glue would be the $0^{++}$ with a mass of $1670 \pm 20$ MeV
\cite{latt_glue}. Experimental searches fave so far found a number of
scalar glueball candidates in the mass region 1300 to 1800 MeV. However,
the interpretation of the data is badly effected by the fact that in
{\em real} QCD, with light quarks, no physical state will be pure glue
-- rather the best one can hope for is an unstable state with only a
small $q \bar q$ component for some (unknown) dynamical reason. We note
that the channel coupling effects induced by decay channels such as $\pi
\pi$ and $K \bar K$ are also quite controversial from the theoretical
point of view. There is clearly room for a great deal of experimental
and theoretical work in this field in the future, with a promise of fame
and fortune for the unambiguous identification of a glueball state.

\section{LIGHT QUARKS}

As we have already explained, the vast majority of theoretical papers
dealing with light quark systems have been based on the constituent
quark model. This approach has achieved a great deal and, as we have
seen in Sect. 2, there is a clear qualitative connection between this model and 
the properties of QCD revealed through lattice calculations.
Nevertheless, the constituent quark model is better suited to dealing
with systems of quarks that are genuinely heavy. Here we take the chiral
properties of QCD very seriously, in order to explore the unique features 
of light quark systems.

\subsection{Trace Anomaly}

If one defines a tensor ${\cal S}^\mu = x_\nu \theta^{\mu \nu}$, with 
$\theta^{\mu \nu}$ the energy-momentum tensor of QCD, one finds
classically that $\partial_\mu {\cal S}^\mu$ is zero. However, if this
were to hold at the quantum level we would find that the nucleon mass
would be zero. Fortunately the divergence of ${\cal S}^\mu$ is no longer
zero in quantum field theory because of the trace anomaly and one finds
for the nucleon mass:
\be
M_N = \langle N | -\frac{9}{4} \alpha_s {\rm Tr} G^2 + m_u \bar u u +
m_d \bar d d + m_s \bar s s | N \rangle.
\label{eq:trace}
\ee
By far the dominant term on the rhs of Eq.(\ref{eq:trace}) is the gluon
trace \cite{xji}. 
(The $u$ and $d$ quark mass terms are known to be of order 40-50
MeV from the sigma commutator and while the $s$-quark mass term is less
well known its not more than 100 MeV or so.)

A full understanding of hadron structure in terms  of QCD
must involve a reasonable physical interpretation of Eq.
(\ref{eq:trace}). For the present it is 
clear that the nucleon mass arises from
non-perturbative gluon interactions and that is certainly reflected
there. There are many famous examples of virial theorems which relate
apparently different physical quantities and it is likely that such a
theorem connects the effective or constituent quark mass appearing in
Eq. (\ref{eq:4}) to the gluon field energy in Eq. (\ref{eq:trace}). 

\subsection{QCD Sum Rules}

The QCD sum rules have had considerable success in relating hadron
masses to various properties of the non-perturbative vacuum. 
The famous Ioffe formula for the nucleon mass \cite{Ioffe}:  
\be
M_N = - \frac{8 \pi^2}{{\cal M}^2} \langle \bar q q \rangle,
\label{eq:ioffe}
\ee
yields a surprisingly accurate value provided the 
Borel mass ${\cal M}$ is taken to be about 1 GeV. It also illustrates
clearly the role of the quark condensate in generating a chiral symmetry
violating property such as mass. On the other hand, the connection to
Eq. (\ref{eq:trace}) is totally unclear. More important, the dependence
on $<\bar q q>$ is incorrect compared with what is found if higher order
condensates are included \cite{DEREK}. In addition, the leading
non-analytic chiral behaviour of the left and right hand sides of Eq.
(\ref{eq:ioffe}) are inconsistent.

\subsection{Quenched Lattice QCD}

Perhaps the most impressive non-perturbative results for the spectrum of
light hadrons has been obtained using quenched lattice QCD. Once the
lattice scale (i.e. the lattice spacing, $a$) is set by fitting one mass
-- either that of the $\rho$ or $\phi$ meson -- the other ground state
meson and baryon masses agree with experiment within about 10\%
\cite{QQCD}. This is a remarkable result, even though one would rather set
$a$ using the string tension, in which case all the masses would be genuine
predictions. On the other hand, the success raises as many questions as
it resolves. Since the chiral properties of quenched QCD (QQCD) are
quite different from those of full QCD \cite{Sharpe}, 
it becomes vital to understand how one can, 
nevertheless, obtain such spectacular agreement. Unfortunately, it will
not be possible to handle realistic light quark masses for sea quarks
(i.e. in quark loops) for quite a few years yet. However, it is known
that chiral symmetry plays a vital role in hadron structure and we have
some clues as to how it might effect lattice calculations.

\subsection{Non-Analytic Behaviour}

We have already seen that spontaneous chiral symmetry breaking in QCD
requires the existence of Goldstone bosons whose masses vanish in the
limit of zero quark mass (the chiral limit).
As a corollary to this, there must be
contributions to hadron properties from Goldstone boson loops.  These
loops have the unique property that they give rise to terms in an
expansion of most hadronic properties as a function of quark mass which
are not analytic.  As a simple example, consider the nucleon mass.  The
most important chiral corrections to $M_N$ come from the processes 
$N \ra N\pi \ra N$ ($\sigma_{NN}$) and $N \ra \Delta \pi \ra N$ 
($\sigma_{N \Delta}$). (We will come to what it means to say these are 
the most important shortly.)  We write
$M_N = M_N^{\rm bare} + \sigma_{NN} + \sigma_{N \Delta}$.
In the heavy baryon limit one has
\be
\sigma_{NN} = - \frac{3 g_A^2}{16 \pi^2 f_\pi^2}
\int_0^\infty dk \frac{k^4 u^2(k)}{k^2 + m_\pi^2}.
\label{eq:3A}
\ee
Here $u(k)$ is a natural high momentum cut-off which is the Fourier
transform of the source of the pion field (e.g. in the cloudy bag model
(CBM) it is $3 j_1(kR)/kR$, with $R$ the bag radius \cite{CBM}). From
the point of view of PCAC it is natural to identify $u(k)$ with the axial
form-factor of the
nucleon, a dipole with mass parameter
$1.02 \pm 0.08$GeV.

Quite independent of the form chosen for the ultra-violet cut-off, one
finds that $\sigma_{NN}$ is a non-analytic function of the quark mass.
The non-analytic piece of $\sigma_{NN}$ is independent
of the form factor and gives
\be
\sigma_{NN}^{LNA} = - \frac{3 g_A^2}{32 \pi f_\pi^2} m_\pi^3
\sim \bar{m}^{\frac{3}{2}}.
\label{eq:4A}
\ee
This has a branch point, as a function of $\bar{m}$, at
$\bar{m} = 0$. Such terms can only arise from Goldstone boson loops.

\subsection{Chiral Extrapolations of Lattice Data}

It is natural to ask how significant this non-analytic behaviour 
is in practice.  If
the pion mass is given in GeV, $\sigma_{NN}^{LNA} = -5.6 m_\pi^3$ 
and at the physical pion mass it is just -17 MeV.
However, at only three times the physical pion mass, $m_\pi = 420$MeV, it
is -460MeV -- half the mass of the nucleon.  If one's aim is to extract
physical nucleon properties from lattice QCD calculations this is
extremely important.  The most sophisticated lattice calculations with
dynamical fermions are only just becoming feasible at such low masses
and to connect to the physical world one must extrapolate from  
$m_\pi \sim 500$MeV to $m_\pi = 140$MeV.
Clearly one must have control of the chiral behaviour.

\begin{figure}[htb]
\centering{\
\rotate{\epsfig{file=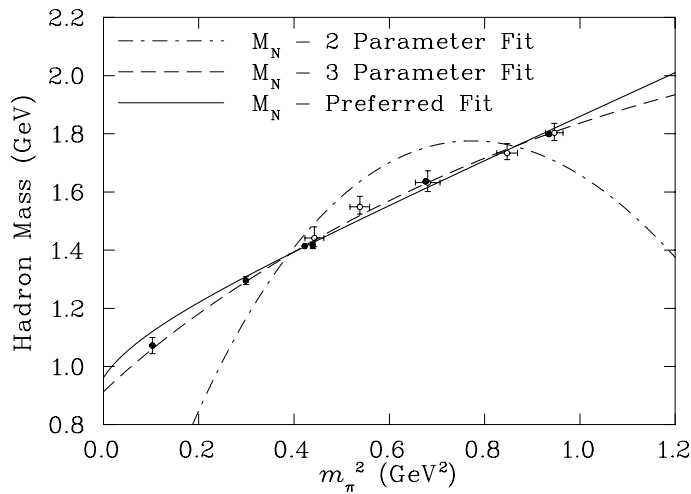,height=9cm}}
\caption{A comparison between phenomenological fitting functions for
the mass of the nucleon -- from Ref. \protect\cite{LEIN}. 
The two parameter fit corresponds to using
Eq.(\ref{eq:5}) with $\gamma$ set equal to
the value known from $\chi$PT.  The three
parameter fit corresponds to letting $\gamma $
vary as an unconstrained fit parameter. The solid line is the 
two parameter fit based on 
the functional form of Eq.(\ref{eq:6}).
\label{fig:FIG2}}}
\end{figure}
{}Figure \ref{fig:FIG2} shows recent 
lattice calculations of $M_N$ as a function of
$m_\pi^2$ from CP-PACS and UKQCD \cite{latt}.  
The dashed line indicates a fit which
naively respects the presence of a LNA term, 
\be
M_N = \alpha + \beta m_\pi^2 + \gamma m_\pi^3,
\label{eq:5}
\ee
with $\alpha, \beta$ and $\gamma$  fitted to
the data.  While this gives a very good fit to the data, the chiral
coefficient $\gamma$ is only -0.761, compared with the value -5.60 
required by chiral
symmetry.  If one insists that $\gamma$ be consistent with QCD the best fit
one can obtain with this form is the dash-dot curve.  This is clearly
unacceptable.

An alternative suggested recently by 
Leinweber et al. \cite{LEIN}, which also
involves just three parameters, is to evaluate $\sigma_{NN}$ and  
$\sigma_{N\Delta}$ with the same
ultra-violet form factor, with mass parameter $\Lambda$, and to fit 
$M_N$ as
\be 
M_N = \alpha + \beta m_\pi^2 + \sigma_{NN}(m_\pi,\Lambda) + 
\sigma_{N\Delta}(m_\pi,\Lambda).
\label{eq:6}
\ee
Using a sharp cut-off ($u(k) = \theta(\Lambda - k)$) these 
authors were able to obtain
analytic expressions for $\sigma_{NN}$ and $\sigma_{N\Delta}$
which reveal the correct LNA
behaviour -- and next to leading (NLNA) in the $\Delta \pi$ case,
$\sigma_{N\Delta}^{NLNA} \sim 
m_\pi^4 \ln m_\pi$.
These expressions also reveal a branch point at $m_\pi = M_\Delta - M_N$, 
which is important if one is extrapolating from large values of $m_\pi$
to the physical value.  The solid curve in Fig. 2 is a two parameter fit
to the lattice data using Eq.(\ref{eq:6}), but fixing $\Lambda$
at a value suggested by CBM simulations to be 
equivalent to the prefered 1 GeV dipole. A small increase in $\Lambda$
is necessary to fit the lowest mass data point, at $m_\pi^2 \sim 
0.1$ GeV$^2$, but clearly one can describe the data very well while
preserving the exact LNA and NLNA behaviour of QCD.

\subsection{The Sigma Commutator}

The analysis of the lattice data for $M_N$, incorporating the correct
non-analytic behaviour, can yield interesting new information concerning
the sigma commutator of the nucleon:
\be
\sigma_N = \frac{1}{3} \langle N| [Q_{i 5},[Q_{i 5},H_{QCD}]] |N\rangle
= \langle N| \bar{m} (\bar{u} u + \bar{d} d) |N\rangle.
\label{eq:8}
\ee
This is a direct measure of chiral SU(2) symmetry breaking in QCD, and
the widely accepted experimental value is 
$45 \pm 8$MeV \cite{SIG_EX}. (Although there
are recent suggestions that it might be as much as 20 MeV 
larger \cite{Kneckt}.)
Using the Feynman-Hellmann theorem one can also write
\be
\sigma_N = \bar{m} \frac{\partial M_N}{\partial \bar{m}} 
= m_\pi^2 \frac{\partial M_N}{\partial m_{\pi}^2}.
\label{eq:9}
\ee
Historically, lattice calculations have evaluated 
$<N| (\bar{u} u + \bar{d} d) |N>$ at large quark mass
and extrapolated this
scale dependent quantity to the ``physical'' quark mass, which had to
be determined in a separate calculation.  The latest result with
dynamical fermions, $\sigma_N = 18 \pm 5$ MeV \cite{SESAM}, 
illustrates how difficult this procedure is. On the other hand, if one
has a fit to $M_N$ as a function of $m_\pi$ which is
consistent with chiral symmetry, one can evaluate $\sigma_N$
directly using Eq.(\ref{eq:9}). Using Eq.(\ref{eq:6}) with a sharp
cut-off yields $\sigma_N \sim 55$ MeV, while a dipole form gives 
$\sigma_N \sim 45$ MeV \cite{SIGMA}. The residual model dependence can only be
removed by more accurate lattice data at low $m_\pi^2$. Nevertheless,
the result $\sigma_N \in (45,55)$ MeV is in very good agreement with the
data.  In
contrast, the simple cubic fit, with $\gamma$ inconsistent with chiral
constraints, gives $ \sim 30$ MeV. Until the experimental situation
regarding $\sigma_N$ improves, it is not possible to draw definite
conclusions regarding the strangeness content of the 
nucleon \cite{Ellis} from this
analysis, but the fact that two-flavour QCD reproduces the current
prefered value should certainly stimulate more work.

\subsection{A Caution}
A number of chiral constituent quark models are in common use, which
involve a Hamiltonian with a pairwise, static one-pion-exchange force
between quarks.  It is important to realise that such models cannot be
consistent with the general chiral constraints of QCD.  
By including only pairwise
pion-exchange interactions, one is omitting Goldstone loops in which the
pion is emitted and absorbed by the same quark \cite{KREIN}.
It is a simple exercise to convince oneself 
that all diagrams, including any number of Goldstone 
boson loops, gluon exchanges and so on, provided they can be cut on a
single pion line, can be exactly written as $\sum_B \sigma_{NB}$.  
Like $\sigma_{NN}$ and $\sigma_{N\Delta}$, $\sigma_{NB}$
describes the process $N \ra B \pi \ra N$, with a 
renormalized $\pi NB$ vertex at each end of
the pion loop.

The LNA behaviour of $M_N$ comes from $B=N$, because only this is degenerate
with the initial nucleon.  The NLNA behaviour comes from $B=\Delta$,
because this is the lowest mass baryon excitation available.
$\sigma_{NN}$ and $\sigma_{N\Delta}$ are
the most important chiral corrections because they produce the LNA and
NLNA behaviour.  Heavier intermediate states will produce only
relatively slow variations of $M_N$ with $m_\pi$ and can be summed
phenomenologically into $\alpha$ and $\beta$ in Eq.(\ref{eq:5}).  
This was the procedure
advocated in the CBM \cite{CBM} and we see that it is
completely consistent with the chiral behaviour of QCD and chiral
perturbation theory, in particular.

We stress that in order to ensure the correct chiral behaviour of hadron
properties one must include all Goldstone loops, including those where it is
emitted and
absorbed by the same quark. 
While it is tempting to argue that
the boson loop on the same quark could be included in the
constituent quark mass, this is incorrect.  Whether one gets LNA or
NLNA behaviour depends on the environment in which the quark finds
itself.  For the nucleon case, if the intermediate
quark plus its spectators has $J=\frac{3}{2}$ 
it is NLNA, while if the three quarks form
a nucleon it is LNA. {\em There is no alternative but to consistently evaluate
Goldstone boson loops at the hadronic level.}

\subsection{Baryon Electromagnetic Properties}
\begin{figure}[t]
\centering{\
\epsfig{file=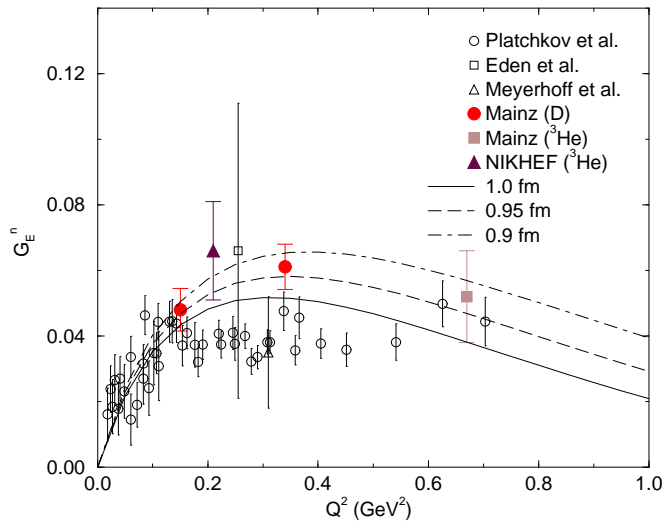,height=8cm}
\caption{Recent data for the neutron electric form factor 
in comparison with CBM calculations for a
confining radius around 0.95fm -- from Ref.\ \protect\cite{LU}.}
\label{fig:neutron}}
\end{figure}
It is a completely general consequence of quantum mechanics that the
long-range charge structure of the proton comes from its $\pi^+$ cloud 
($p \ra n \pi^+$),
while for the neutron it comes from its $\pi^-$ cloud ($n \ra p \pi^-$). 
However it is not
often realized that the LNA contribution to the nucleon charge radius
goes like $\ln m_\pi$ and diverges as $\bar{m} \ra 0$ \cite{LN}.  
This can never be described by a
constituent quark model. Figure \ref{fig:neutron}
shows the latest data from Mainz and Nikhef
for the neutron electric form factor, in comparison with CBM
calculations for a confinement radius between 0.9 and 1.0 fm. The
long-range $\pi^-$ tail of the neutron plays a crucial role.

The situation for baryon magnetic moments is also very interesting.  
The LNA contribution in this case arises from the diagram where 
the photon couples to the pion loop.  As this
involves two pion propagators the expansion of the proton and neutron
moments is: 
\be
\mu^{p(n)} = \mu^{p(n)}_0 \mp \alpha m_\pi + {\cal O}(m_\pi^2).
\label{eq:10}
\ee
Here $\mu^{p(n)}_0$ is the value in the chiral limit and the
linear term in $m_\pi$ is proportional to $\bar{m}^{\frac{1}{2}}$, 
a branch point at $\bar{m} = 0$.  The
coefficient of the LNA term is $\alpha = 4.4 \mu_N-$GeV$^{-1}$. 
At the physical pion mass this LNA
contribution is $0.6\mu_N$, which is almost a third of the neutron magnetic
moment. {\em No constituent quark model can or should get better agreement
with data than this.}

Just as for $M_N$, the chiral behaviour of $\mu^{p(n)}$ is vital to a correct
extrapolation of lattice data. One can obtain a very satisfactory fit to
some rather old data, which happens to be the best available,
using the simple Pad\'e \cite{MAGMOM}:
\be
\mu^{p(n)} = \frac{\mu^{p(n)}_0}{1 \pm \frac{\alpha}{\mu^{p(n)}_0} m_\pi +
\beta m_\pi^2}
\label{eq:11}
\ee
The data can only determine two parameters and Eq.(\ref{eq:11})
has just two free parameters while guaranteeing the correct LNA
behaviour as $m_\pi \ra 0$ {\bf and} the correct behaviour of HQET 
at large $m_\pi^2$.  The
extrapolated values of $\mu^p$ and $\mu^n$ 
at the physical pion mass, $2.85 \pm 0.22 \mu_N$ and $-1.90 \pm 0.15
\mu_N$ are currently the best estimates from non-perturbative QCD
\cite{MAGMOM}. For more details of this fit and the application 
of similar ideas to other members of the
nucleon octet, as well as the strangenesss magnetic moment of the
nucleon, we refer to the presentation of D. Leinweber at this conference
\cite{Derek_fb16}.

\subsection{Deep-Inelastic Scattering}

Although deep inelastic scattering has predominantly been used as a
testing ground for perturbative QCD, the parton distributions contain
direct information on the energy-momentum distribution of quarks and
gluons inside the hadron. This is vital to our understanding of 
how hadron structure is realized in non-perturbative
QCD. Of course, this information is viewed in a particular frame of
reference, the infinite momentum frame.
The problem of zero modes on in light-cone field theory is widely
appreciated and the chiral behaviour of hadronic properties is usually
believed to be part of that general problem.  However, the LNA behaviour
of these properties can give us some insight.

The observation that  
non-analytic behaviour in $\bar{m}$ can only come from Goldstone boson
loops is just as true on the light-cone as anywhere else.  Consider the
process $N \ra N \pi \ra N$ in light-cone field theory.
It is a straightforward exercise to show that the expression
obtained using the Feynman rules for light-cone field theory
does indeed have the LNA behaviour required by QCD.
This seems natural, why bother to mention it?  The point is that
this LNA behaviour can only come from a $\pi N$ intermediate
state.  The subtle correlations in the pion which guarantee that
$m_\pi^2 \propto \bar{m}$,
vanishing in the chiral limit, must be included.  One will {\em never}
obtain
the LNA behaviour by including a few or even quite a lot of leading Fock
components in the nucleon wavefunction.  

This simple argument leads us to conclude that both $N \pi$
and $\Delta \pi$ Fock
components must be incorporated in the light-cone wave function of the
nucleon if one is satisfy chiral constraints.  One now famous
consequence of this is the excess of $\bar{d}$ over $\bar{u}$
quarks in the proton,
predicted \cite{Thomas83} in 1983
and observed by NMC in 1990 \cite{NMC}.  This measurement, and
the important developments since then \cite{LATER},
have given us
vital new information about the non-perturbative chiral structure of the
nucleon.  In this light it satisfying to see the remarkable result
recently obtained by Melnitchouk et al. \cite{MST2000},
which relates the LNA behaviour
of the excess of $\bar{d}$ over $\bar{u}$ quarks to
the LNA behaviour of the nucleon wave function renormalization constant:
\be
\int_0^1 dx \left( \bar{d}(x) - \bar{u}(x) \right) |_{LNA}
\sim m_\pi^2 \ln m_\pi .
\label{eq:12}
\ee

\section{Future Challenges in Spectroscopy}

The calculation of the properties of highly excited baryons within
lattice QCD is a very difficult challenge. However, for the lowest
negative parity states new techniques have been developed which seem
very promising \cite{negparity}. On the experimental side we can expect
a tremendous wealth of new data from Jefferson Lab in the next few
years. This should greatly clarify the situation regarding ``missing
states'' and give us a much clearer picture concerning many other
resonances. 

One of the major theoretical challenges for the near future is the 
need to deal with the coupling of hadron resonances to various
meson-baryon channels (open and closed). Not only do we have a problem that
some states expected in the quark model are missing but there may be
other states that should not be considered quark states at all. A famous
example is the $\Lambda(1405)$, which almost certainly results from the
extremely strong attraction in the $\Sigma \pi$ s-wave, coupled to $\bar
K N$ \cite{L1405}. Another candidate which has attracted a great deal of
theoretical attention is the Roper. The most recent study by the
J\"ulich group, including the effect of coupled $\pi N, \pi \Delta$ and
$\sigma N$ channels, strongly suggests that the Roper is not a quark
model state \cite{speth}. 
This would certainly resolve a number of problems with its
low mass.

While these interpretations of the $\Lambda(1405)$ and the Roper seem to
be correct, there is a real danger in an uncontrolled coupled channels
approach to such problems. As illustrated by the classic case of the
Chew-Low model of the $\Delta(1232)$, 
it is always possible to generate a resonance
through multiple scattering. On the other hand, we know that in the
absence of open channels QCD predicts the existence of hadrons. It is
essential in unravelling the nature of observed resonance states
that one must use a consistent model of the internal structure of the 
hadrons involved in order to calculate the appropriate coupling to
relevant meson-baryon channels. The classic CBM work on the
$\Delta(1232)$ showed that in such an approach it was clear that it is
primarily a three-quark state, not the result of strong pion-nucleon
rescattering. The J\"ulich group finds a similar result for the
$S_{11}(1535)$ -- even though its coupling to $\eta N$ is very strong. 
{}Finally, we note that while the examples quoted have been
baryon resonances, exactly the same questions arise for mesons
\cite{oset}. This is a field of study which is just beginning in earnest.

\section{Conclusion}

It is clear that there have been some outstanding developments in our
understanding of hadron structure in terms of QCD over the last 25
years. Yet the next 10 promise much more. We can expect data on 
hadron electroweak form factors of increasing precision over a vastly wider 
range of kinematic variables. We can also expect an entirely new range
of observables, including for the nucleon: virtual Compton
scattering (VCS), deeply VCS, spin-dependent Compton scattering,
transition form factors to (and between)  
new baryon resonances, determinations of
$G_E^s$ and $G_M^s$, the nucleon anapole moment and semi-inclusive
deep-inelastic scattering data.

On the theoretical side there will be rapid progress in lattice QCD,
with improved actions and faster computers making it possible to include
dynamical quarks with masses approaching the physical region. Improved
interaction between phenomenology and lattice simulations will lead to
ever more reliable chiral extrapolations of the physical properties of
hadrons. There will also be serious and consistent studies of the
effects of channel coupling, both for regular meson and baryon
resonances but also for hybrids and glueballs. Finally, we can expect a
great deal more productive feedback between the results of QCD inspired
models and lattice simulations.

\begin{center}
{\bf ACKNOWLEDGEMENTS}
\end{center}
It is a pleasure to acknowledge the collaboration and helpful
discussions with many staff, students and visitors at the CSSM who have
contributed to my understanding of the problems discussed here,
especially C. Boros, G. Krein, D. Leinweber, D. Lu, W. Melnitchouk, F. Steffens,
K. Tsushima, A. Williams and S. Wright.
This work was supported by the Australian Research Council and the
University of Adelaide.


\begin{thebibliography}{9}
%
\bibitem{cotanch}
F.~J.~Llanes-Estrada and S.~R.~Cotanch,
Phys.\ Rev.\ Lett.\  {\bf 84} (2000) 1102.
%
\bibitem{Toki}
Y.~Koma, H.~Suganuma, K.~Amemiya, M.~Fukushima and H.~Toki,
hep-ph/9912347.
%
\bibitem{bernard}
C.~Bernard {\it et al.},
hep-lat/0002028.
%
\bibitem{gluonprop}
{}F.~D.~Bonnet {\it et al.},
hep-lat/0002020.
%
\bibitem{alkofer}
R.~Alkofer and L.~von Smekal,
to appear in Nucl. Phys., hep-ph/0004141.
%
\bibitem{quark}
J.~I.~Skullerud and A.~G.~Williams,
Nucl.\ Phys.\ Proc.\ Suppl.\  {\bf 83-84} (2000) 209.
%
\bibitem{aoki}
S.~Aoki {\it et al.}  [JLQCD Collaboration],
Phys.\ Rev.\ Lett.\  {\bf 82} (1999) 4392.
%
\bibitem{roberts}
F.~T.~Hawes {\it et al.},
Phys.\ Lett.\  {\bf B440} (1998) 353
[nucl-th/9807056].
%
\bibitem{SDE}
C.~D.~Roberts and A.~G.~Williams,
Prog.\ Part.\ Nucl.\ Phys.\  {\bf 33} (1994) 477.
%
\bibitem{IK_missing}
N.~Isgur and G.~Karl,
Phys.\ Rev.\  {\bf D18} (1978) 4187.
%
\bibitem{capstick}
S.~Capstick and N.~Isgur,
Phys.\ Rev.\  {\bf D34} (1986) 2809.
%
\bibitem{Gloz1}
L.~Y.~Glozman and D.~O.~Riska,
Phys.\ Rept.\  {\bf 268} (1996) 263
[hep-ph/9505422].
%
\bibitem{nathan}
N.~Isgur,
nucl-th/9908028.
%
\bibitem{Gloz2}
L.~Y.~Glozman,
nucl-th/9909021.
%
\bibitem{PRL81_5760}
G.~S.~Adams {\it et al.}  [E852 Collaboration],
Phys.\ Rev.\ Lett.\  {\bf 81} (1998) 5760.
%
\bibitem{Hyb_latt}
P.~Lacock {\it et al.} [UKQCD Collaboration],
Phys.\ Rev.\  {\bf D54} (1996) 6997.
%
\bibitem{latt_glue}
C.~Michael,
Nucl.\ Phys.\  {\bf A655} (1999) 12
[hep-ph/9810415].
%
\bibitem{xji}
X.~Ji,
Phys.\ Rev.\ Lett.\  {\bf 74} (1995) 1071
[hep-ph/9410274].
%
\bibitem{Ioffe}
B.\ Ioffe, Nucl.\ Phys.\ {\bf 188} (1978) 317
%
\bibitem{DEREK}
D.~B.~Leinweber,
Annals Phys.\  {\bf 254} (1997) 328
[nucl-th/9510051].
%
\bibitem{QQCD}
S.~Aoki {\it et al.}  [CP-PACS Collaboration],
Phys.\ Rev.\ Lett.\  {\bf 84} (2000) 238.
%
\bibitem{Sharpe}
J.~N.~Labrenz and S.~R.~Sharpe,
Phys.\ Rev.\  {\bf D54} (1996) 4595
[hep-lat/9605034].
%
\bibitem{CBM}
S.~Theberge, A.~W.~Thomas and G.~A.~Miller,
Phys.\ Rev.\  {\bf D22} (1980) 2838;
%
A.~W.~Thomas,
Adv.\ Nucl.\ Phys.\  {\bf 13} (1984) 1.
%
\bibitem{latt}
S.~Aoki {\it et al.}  [CP-PACS-Collaboration],
Phys.\ Rev.\  {\bf D60} (1999) 114508
;
%
C.~R.~Allton {\it et al.}  [UKQCD Collaboration],
Phys.\ Rev.\  {\bf D60} (1999) 034507
.
%
\bibitem{LEIN}
D.~B.~Leinweber {\it et al.},
Phys.\ Rev.\  {\bf D61} (2000) 074502
[hep-lat/9906027].
%
\bibitem{SIG_EX}
J.~Gasser, H.~Leutwyler and M.~E.~Sainio,
{\em Phys. Lett.} {\bf B253}, 252 (1991).
%
\bibitem{Kneckt}
M.~Knecht,
{\tt hep-ph/9912443}.
%
\bibitem{SESAM}
{\bf SESAM} Collaboration, S.~Gusken {\it et al.},
{\em Phys.\ Rev.}\  {\bf D59}, 054504 (1999).
%
\bibitem{SIGMA}
D.~B.~Leinweber {\it et al.},
Phys.\ Lett.\  {\bf B482} (2000) 109
[hep-lat/0001007].
%
\bibitem{Ellis}
J.~Ellis,
in Proc. this conference, 
hep-ph/0005322.
%
\bibitem{KREIN}
A.~W.~Thomas and G.~Krein,
Phys.\ Lett.\  {\bf B456} (1999) 5
[nucl-th/9902013].
%
\bibitem{LU}
D.\ H.\ Lu {\it et al.}, in Proc. this conference;
D.~H.~Lu {\it et al.},
Phys.\ Rev.\  {\bf C60} (1999) 068201
[nucl-th/9807074].
%
\bibitem{LN}
D.~B.~Leinweber and T.~D.~Cohen,
Phys.\ Rev.\  {\bf D47} (1993) 2147
[hep-lat/9211058].
%
\bibitem{MAGMOM}
D.~B.~Leinweber {\it et al.}, 
Phys.\ Rev.\  {\bf D60} (1999) 034014
[hep-lat/9810005].
%
\bibitem{Derek_fb16}
D.~B.~Leinweber and A.~W.~Thomas, in Proc. this conference.
%
\bibitem{Thomas83}
A.~W.~Thomas,
Phys.\ Lett.\  {\bf B126} (1983) 97.
%
\bibitem{NMC}
P.~Amaudruz {\it et al.}  [New Muon Collaboration],
Phys.\ Rev.\ Lett.\  {\bf 66} (1991) 2712.
%
\bibitem{LATER}
E.~A.~Hawker {\it et al.}  [NuSea Collaboration],
Phys.\ Rev.\ Lett.\  {\bf 80} (1998) 3715.
%
\bibitem{MST2000}
A.~W.~Thomas, W.~Melnitchouk and F.~M.~Steffens,
hep-ph/0005043.
%
\bibitem{negparity}
F.~X.~Lee and D.~B.~Leinweber,
Nucl.\ Phys.\ Proc.\ Suppl.\  {\bf 73} (1999) 258.
%
\bibitem{L1405}
N.~Kaiser, P.~B.~Siegel and W.~Weise,
Nucl.\ Phys.\  {\bf A594} (1995) 325
[nucl-th/9505043];
%
E.~A.~Veit {\it et al.},
Phys.\ Rev.\  {\bf D31} (1985) 1033;
%
R.~H.~Dalitz and A.~Deloff,
J.\ Phys.\ G {\bf G17} (1991) 289.
%
\bibitem{speth}
C.~Schutz {\it et al.},
Phys.\ Rev.\  {\bf C57} (1998) 1464.
%
\bibitem{oset}
D.~Cabrera, E.~Oset and M.~J.~Vicente-Vacas,
nucl-th/0006029.
%
\end{thebibliography}
\end{document}